\documentclass[preprint2]{aastex}
\newcommand{\mperm}{\mu_0}
\newcommand{\mdiff}{\eta}
\newcommand{\cmnt}[1]{}
\newcommand{\comm}[1]{}
\newcommand{\ignore}[1]{}

\def\gsim{\lower.4ex\hbox{$\;\buildrel >\over{\scriptstyle\sim}\;$}}
\def\lsim{\lower.4ex\hbox{$\;\buildrel <\over{\scriptstyle\sim}\;$}}

\def\beg{\begin{eqnarray}}
\def\ende{\end{eqnarray}}
\renewcommand{\vec}[1]{\mbox{\boldmath $#1$}}

\shorttitle{The  columnar gallium Tayler experiment}
\shortauthors{G. R\"udiger et al.}

\begin{document}

\title{Critical fields and growth rates of the Tayler instability as probed  
by a  columnar gallium experiment}

\author{G\"unther R\"udiger, Marcus Gellert, Manfred Schultz, Klaus G. Strassmeier}
\affil{Leibniz-Institut f\"ur Astrophysik Potsdam, An der Sternwarte 16,
D-14482 Potsdam, Germany}

\author{Frank Stefani, Thomas Gundrum, Martin Seilmayer, Gunter Gerbeth}
\affil{Helmholtz-Zentrum Dresden-Rossendorf, P.O. Box 510119, D-01314 Dresden,
Germany}

\begin{abstract}
Many astrophysical phenomena (such as the slow rotation of neutron stars or the
rigid rotation of the solar core) can be explained by the action of the Tayler 
instability of toroidal magnetic fields in the radiative zones of stars. In order 
to place the theory of this instability on a safe fundament it has been realized 
in a laboratory  experiment measuring the critical field strength, the growth
rates as well as the shape of the supercritical modes. A strong electrical current 
flows through a liquid-metal confined in a resting columnar container with an 
insulating outer cylinder. As the very small magnetic Prandtl number of the
gallium-indium-tin alloy does not influence the critical Hartmann number of
the field amplitudes, the electric currents for marginal instability can
also be computed with direct numerical simulations. The results of this
theoretical concept are confirmed by the experiment. Also the predicted 
growth rates of the order of minutes for the nonaxisymmetric perturbations are certified
by the measurements. That they do not directly depend on the size of the
experiment is shown as a consequence of the weakness of the applied fields and
the absence of rotation.

\end{abstract}

\keywords{stars: rotation --- stars: magnetic field --- instabilities
--- magnetohydrodynamics}

\section{Introduction}

Sufficiently strong, not current-free toroidal magnetic fields
become unstable due to the non-axisymmetric Tayler instability
(``TI''; Tayler 1957, 1960,  Vandakurov 1972, Tayler 1973). The
necessary energy is provided by the electric current, hence the
instability also exists without plasma motion, e.g. due to
stellar rotation. 
So far, mostly astrophysics-related numerical
simulations of the TI are very rare (Braithwaite \& Spruit 2004,
Braithwaite 2006, Gellert et al. 2008,  Moll et al. 2008).

The existence of the magnetic Tayler instability likely has
immense astrophysical consequences. Ott et al. (2006) 
concluded from their supernova core-collapse simulations (with
the rotation of the initial iron core as the free parameter) that
the rotation period of a newly born neutron star should not exceed
1~ms, in contrast to the observations that peak with periods
around 10--100~ms. Berger et al. (2005) argue for an upper limit
of 10~km/s rotation velocity for white dwarfs using their
spectroscopy of the rotational broadening of the Ca\,{\sc ii}~K
line. Neutron stars as well as white dwarfs are compact remnants
of stellar cores that exhibit a specific angular momentum of
$10^{13\dots 14}$~cm$^2$/s. However, for their progenitors, Berger
et al.'s simulations provide values of more than
10$^{16}$~cm$^2$/s, indicating two missing orders of magnitude
between the hydrodynamic theory and the observations. Suijs et al.
(2008) indeed show that an evolution scenario for stars with
1--3~M$_\sun$ that includes small-scale Maxwell stresses can
explain the extreme spin down of the stellar core by typically two
orders of magnitude.

Furthermore, we know from helioseismology that the solar radiative
core rotates basically rigidly despite that the microscopic
viscosity of the solar plasma yields a diffusion time much longer
than the Sun's lifetime. One would need an effective viscosity of
about $10^5$~cm$^2$/s to explain the decay of an initial rotation
law within the lifetime of the Sun.
It is thus tempting to probe whether an instability-driven angular
momentum transport -- either due to magnetic instabilities of
fossil fields and/or of induced internal toroidal fields -- is
strong enough to produce an increase of the microscopic viscosity
in the radiative zone by several orders of magnitude (Eggenberger
et al. 2005). For example, one finds for the upper part of the
solar radiative core about 600~Gauss as the minimum amplitude of a
toroidal field to become unstable.

\begin{figure}[hbt]
\epsscale{0.95} 
\plotone{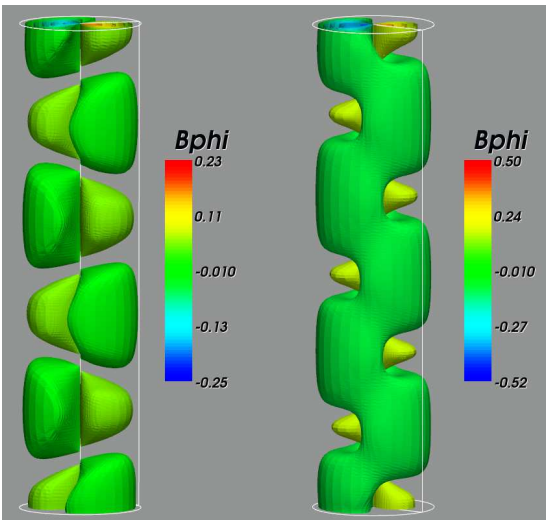} 
\caption{Isosurface
plot of the magnetic field perturbation $b_\phi$ for a magnetic
Prandtl number ${\rm Pm}=0.01$ (left) and ${\rm Pm}=1$ (right).
{ The direct numerical simulation  was carried out with} an insulating outer
cylinder, $R_{\rm in}=0$, { a Hartmann number} $\rm Ha=30$, and
perfect-conducting endplates. Note the rather high amplitudes of
the perturbation patterns (normalized with the outer azimuthal
background field, not shown). } \label{marcus1}
\end{figure}

The lithium at the surface of cool main-sequence stars 
depletes with a timescale of $\approx$1~Gyr. Lithium is burned at
temperatures in excess of $2.6 \cdot 10^6$~K that, in case of
today's Sun, exist about 40,000~km below the base of the
convection zone. Consequently, there must be a diffusion
process between the upper layers and the regions of the burning
temperature. Its time scale must
be one or two orders of magnitude  shorter than the molecular
diffusion  in order to explain the above lithium decay time.
On the other hand, any enhanced chemical mixing comes along with
an intensified mixing of the angular momentum. Considering
transport processes in the radiative interior of massive
(15~M$_\odot$) main-sequence stars, Spruit (2002) and Maeder \&
Meynet (2003, 2005) computed viscosities up to $10^{13}$~cm$^2$/s
for an equatorial rotational velocity of 300~km/s with which the
internal stellar rotation becomes rigid after a few thousand
years. Heger et al. (2005) and Woosley \& Heger (2006) followed on
this basis the rotational history of massive main-sequence stars
until their collapse-end.

Rotationally induced mixing is also included in the stellar
evolution codes by Heger \& Langer (2000), and Yoon et al. (2006)
presented evolutionary models of rotating stars with different
metalicity. As even the lifetime  and thus the evolutionary
path in the H-R diagram is influenced by the mixing, the action
of magnetic instabilities must also influence the ages of young
stellar clusters.  This may alter the rotation-activity-age
relation for low-mass stars which very much depends on the age
determination of young clusters (Barnes 2010). Brott et al.
(2008) demonstrate that the magnetic-induced chemical mixing in
massive stars must even be reduced to avoid conflicts with the
observations.

A dynamo action on the basis of TI, as proposed by Spruit (1999,
2002) and Braithwaite (2006), could not be confirmed so far (Zahn
et al. 2007). The  suggestion is that differential rotation and a
magnetic kink-type instability could jointly drive a dynamo in
stellar radiative zones. If existent, such a dynamo would also be
effective for the angular momentum transport and the chemical
mixing in stellar interiors. However, more detailed numerical
models are needed to investigate the dynamo efficiency of the TI,
e.g. by computing the magnetic-induced (magnetic) dissipation {\em
and} the possible existence of the predicted kind of
$\alpha$-effect.

In general, the search for the  possible existence of kinetic/magnetic
helicity and/or $\alpha$-effect leads to the investigation of the stability
of background fields with finite current helicity $\vec{B} \cdot {\rm
curl}\ \vec{B}$. Such a field geometry results by a combination of axial
currents with axial fields. The reason is that the TI of an purely
azimuthal field does not produce an instability pattern with finite
helicity so that such simple fields would not produce any TI-induced
$\alpha$-effect.
The rotation of the container only plays here a minor role but spontaneous
parity breaking bifurcations of solutions with an equal mixture of both
signs of helicity in MHD system which might be unstable,  is now under discussion (Chatterjee et al.
2011, Bonanno et al. 2012). 

 This situation is drastically changed if an axial field
is added to the current-induced azimuthal field. Small-scale kinetic and
current helicities immediately result which are  anticorrelated to the current helicity of the
background  field. The corresponding $\alpha$-effect is highly anisotropic
($\alpha_{\phi\phi} \cdot \alpha_{zz}  < 0$) where  the $\alpha_{\phi\phi}$
has the same sign as the current helicity of the background field (Gellert et
al. 2011).
On the other hand, it is clear that the
combination of axial currents and axial fields may drastically change
the excitation conditions. Bonanno \& Urpin (2008) found a stabilization
of axisymmetric TI modes for finite axial fields. Also the
nonaxisymmetric modes are stabilized for fields with large  pitch angle
$|B_z/B_\phi|$ while the fastest growing modes always possess higher
azimuthal wave numbers (Bonanno \& Urpin 2011, R\"udiger et al. 2011b).

{ To our knowledge, the pure form of the TI,
i.e. its action in an incompressible fluid (Tayler 1960),
was unknown in experimental physics so far (see also Meynet
\& Maeder 2005). This is in contrast to the comprehensive
experience in plasma physics where the compressible counterpart
of the TI is better known as the kink instability ($m=\pm 1$) in a z-pinch,
i.e. the limit of the Kruskal-Shafranov instability when the
safety factor goes to zero (Bergerson et al.  2006).              
In order to place the many magnetic numerical simulations in       
stellar physics onto a safe fundament, we conducted and present  
here the first successful realization of the TI in an experiment 
with an incompressible fluid. 
The critical magnetic field strength and the supercritical growth 
rates are calculated both quasi-analytically and by numerical 
simulations and are compared with the onset of the instability in 
the columnar ``gallium Tayler instability experiment'', dubbed {\sl GATE}. 
The design and the  technical details of the experiment are given by 
Seilmayer et al. (2012).

\section{Theory and simulations}

A cylindrical Taylor-Couette container is considered confining a
toroidal magnetic field of given amplitude.  The container
possesses an inner and an outer cylinder with radii $R_{\rm in}$
and $R_{\rm out}$ with $r_{\rm in}={R_{\rm{in}}}/{R_{\rm{out}}}$.
The radius of the inner cylinder is assumed as very small including 
the limit zero. The fluid  between the
cylinders is assumed to be incompressible with uniform density and
dissipative with both the kinematic viscosity $\nu$ and the
magnetic diffusivity $\mdiff$.

The solution of the stationary induction equation inside the outer
cylinder under the presence of a uniform electric current $I$
reads 
\beg B_\phi=\frac{I}{5 R^2_{\rm out}} R \label{basic}
\ende
(see Roberts 1956; Pitts \& Tayler 1985; Spies 1988).  R\"udiger \& Schultz
(2010) have shown for all azimuthal Fourier modes that for resting
cylinders the critical value of the Hartmann number
\beg
    {\rm Ha} = \frac{B_{\rm out} R_{\rm out} }{\sqrt{\mperm \rho \nu \mdiff}}.
 \label{Hart}
 \ende
does {\em not} depend on the magnetic Prandtl number $\rm
Pm=\nu/\mdiff$. It is thus allowed for the calculation of the
critical Hartmann number to solve the complete set of equations
only for the numerically most simple case of $\rm Pm=1$. This is
not true, however, with respect to the growth rates.

If the radial profiles of the azimuthal field are not too steep
the current-driven TI is mainly a non-axisymmetric instability.
While for the stability of non-axisymmetric modes the necessary
and sufficient condition is
\beg \frac{{\rm{d}}}{{\rm{d}}R}( R
B_\phi^2) < 0 \label{cond1}
\ende
(Tayler 1973), the same reads for axisymmetric modes as
\beg
\frac{{\rm{d}}}{{\rm{d}}R}\left( \frac{B_\phi}{R} \right)^2 < 0
\label{cond2}
\ende
(Velikhov 1959, Chandrasekhar 1961). The {\em stable} domain for
axisymmetric modes thus always includes the profile 
$B_\phi\propto R$ noted  in Eq. (\ref{basic}) which after 
Eq. (\ref{cond1}) will be unstable against  non-axisymmetric perturbations. 
In this case both the modes with $+m$ and $-m$ are simultaneously
excited so that the resulting field and flow components do not
perform any azimuthal drift. The most
unstable mode is $|m|=1$. Figure~\ref{marcus1} shows an
isosurface plot of the nonaxisymmetric field pattern of the
dynamic azimuthal component $b_\phi$ from nonlinear simulations
for both $\rm Pm=0.01$ and $\rm Pm=1$. The field is normalized
with the fixed outer value of the applied toroidal field. One
finds a maximum perturbation amplitude of around 50\% of the
applied field for $\rm Pm=1$. This surprisingly high value varies
only weakly with $\rm Pm$: it only varies by a factor of two if
$\rm Pm$ varies by two orders of magnitude. The wave number is
also nearly constant.

The nonlinear MHD code of Gellert et al. (2007) can  be used to
compute the critical Hartmann number for $\rm Pm=1$, which then
holds for all $\rm Pm$. A first calculation concerns an infinite
cylinder with insulating walls while a second one concerns a
closed cylinder with perfect-conducting endplates. The resulting
instability conditions are ${\rm Ha} \simeq 22.3$ for the infinite
container and ${\rm Ha} \simeq 22.5$ for the closed container. As
expected for closed containers the TI is slightly stabilized by
the endplates because of the wave number restrictions.

The  Hartmann number (Eq.~\ref{Hart}) determines the electrical
current via
\beg I= 5 {\rm Ha} \sqrt{\mu_0\rho\nu\eta},
\label{curr}
\ende
where for the material constant of the used gallium-indium-tin
alloy one finds  $\sqrt{\mu_0\rho\nu\eta}= 25.3$ in c.g.s. for a
temperature of 300 K  ($\rho= 6.44$\ g/cm$^3$, $\eta=2440$\
cm$^2$/s and $\nu=3.25\cdot 10^{-3}$\ cm$^2$/s). With these
numbers the  minimum electric current for excitation of TI becomes
2.8~kA. This is a minimum  value which cannot be optimized by variation of 
the radial size  of the container.

\section{The linearized system of equations}
Sofar the magnetic Prandtl number does not play any role. However,
this is not true for both the growth rates and also the wave
numbers of the unstable modes for supercritical magnetic fields.
Because of the low magnetic Prandtl number for liquid
metals they can accurately be computed only with a linear code.
The linearized MHD equations for the flow and field perturbations
in a conducting fluid subject to a magnetic background field
$\vec{B}$ are
\begin{eqnarray}
\lefteqn{\frac{\partial \vec{u}}{\partial t}=-\nabla(p/\rho)+}\nonumber\\
&& \quad\quad\quad + \nu \Delta \vec{u} + {\rm curl} {\vec{B}}\times \vec{b}+ {\rm curl} \vec{b}\times {\vec{B},}\nonumber\\
\lefteqn{\frac{\partial \vec{b}}{\partial t}= \eta \Delta \vec{b}+ {\rm curl}\big(\vec{u}\times {\vec{B}}\big)}
\label{mhd}
\end{eqnarray}
with ${\rm div}\ {\vec{u}} =  {\rm div}\ {\vec{b}} = 0$ and  $p$
the pressure fluctuations. The boundary conditions for the flow
are always no-slip and the outer cylinder is always assumed to be
insulating. For the space within the inner cylinder, two magnetic
conditions are possible: a vacuum or a perfect conductor. None of
them exactly fits in the limit $R_{\rm in}\to 0$ to the regularity
conditions at the axis which for $m=1$ reads ${\rm d} b_R/{\rm
d}R= {\rm d} b_\phi/{\rm d}R = b_z=0$. These different boundary
conditions are applied at $R_{\rm{in}}$. In a series of
calculations the limit $R_{\rm{in}}\to 0$ is approached. We found
highly convergent solutions for the inner perfect-conductor
solutions while the vacuum condition creates severe numerical
problems for $r_{\rm in}\to 0$.

The wave number is varied as long as the corresponding Hartmann
number takes its minimum. A vanishing growth rate provides the
critical Hartmann number or, using Eq. (\ref{curr}), the critical
electric current of the TI. Figure~\ref{H} provides the behavior
of $\rm Ha$ for $R_{\rm in}\to 0$.  In this case one finds that
$\rm Ha$ is almost independent of $r_{\rm in}$ but for intermediate
values $\rm Ha$ increases. For very small $R_{\rm in}$ the result
is $\rm Ha=22$. The critical strength of the electric current is
thus 2.78~kA, corresponding to Eq.~(\ref{basic}) to an outer
azimuthal field of about 110~Gauss. Evidently, the above results
of the nonlinear simulations for ${\rm Pm}=1$ and the limit of the
linear calculations for $R_{\rm in}\to 0$ with ${\rm Pm}=10^{-6}$
indeed coincide.

In a resting container and a fluid with magnetic Prandtl number of order unity
one  has only  two frequencies: the diffusion frequency $\omega_\eta=\eta/R_{\rm out}^2$ 
and the  Alfv\'en frequency $\Omega_{\rm A}=B_\phi/\sqrt{\mu_0\rho R^2_{\rm out}}$ 
of the toroidal field. Its ratio defines the limit of low-conductivity
(or weak-field, $\omega_\eta\gg \Omega_{\rm A}$) and high-conductivity 
(or strong-field, $\omega_\eta\ll \Omega_{\rm A}$). Hence, the growth 
rate of the instability must be a linear function of these two frequencies 
in form of $\Omega_{\rm A}$ and $(\Omega_{\rm A}/\omega_\eta)\Omega_{\rm A}$. 
It is also clear that in the high-conductivity limit only the term linear in 
$\Omega_{\rm A}$ can appear. In the low-conductivity limit the form quadratical 
in $\Omega_{\rm A}$, i.e. 
\beg
\omega_{\rm gr} = \Gamma \frac{B^2_{\rm out}}{\mu_0\rho\eta}   ,
\label{omg}
\ende
may dominate where the factor $\Gamma$ for wide gaps varies only by a factor of
four when the magnetic Prandtl number varies by four orders of
magnitude  (R\"udiger et al. 2011a). Note that in the low-conductivity limit
the radial scale does not occur in Eq.~(\ref{omg}).  For highly
supercritical Hartmann numbers Eq. (\ref{omg}) smoothly changes to the linear
relation $\omega_{\rm gr}\propto \Omega_{\rm A}$ which does no longer depend on
the microscopic diffusivities but which strongly decreases for increasing
radius. The transition to the linear relation requires the higher Hartmann
numbers the smaller the magnetic Prandtl number is. To date the transition can 
only be seen numerically, not experimentally. 

\begin{figure}[hbt]
\epsscale{1.0} 
\plotone{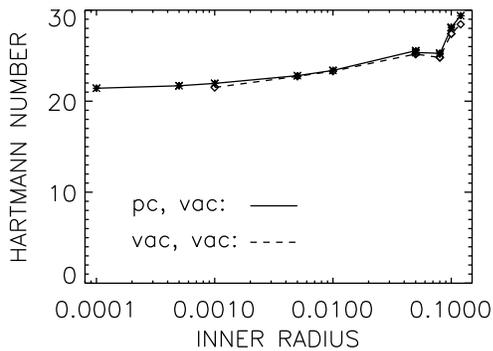} 
\caption{Numerical values of the
critical Hartmann number ${\rm Ha}$ for large gaps ($r_{\rm in}\ll
1$) between (infinitely) long cylinders. The outer boundary
condition is vacuum, within the inner cylinder is vacuum or a
perfectly conducting material. The curves are valid for all $\rm
Pm$.} \label{H}
\end{figure}

For $\rm Pm\simeq 10^{-(5\dots 6)}$ the calculation of the growth
rates leads to $\Gamma\simeq 0.04 $. The high value of the magnetic
diffusivity of the  fluid conductor leads to growth times of about
300~s for supercritical electric currents of order 4~kA
(Fig.~\ref{results}). These time scales are so long that a
secondary circulation may arise in the experiment due to the internal
Joule heating of the fluid. In the linear theory there is no magnetic
stabilization mechanism as, e.g., it exists for strong fields for
 the magnetorotational instability (Balbus \& Hawley 1991).

The optimized wave numbers prove to be nearly independent of the magnetic field  
and the geometry of the container (Fig. \ref{wavenumber}). They describe the 
axial scale $\delta z$ of the instability pattern after the relation 
\beg 
\frac{\delta z}{R_{\rm out}-R_{\rm in}}=\frac{\pi}{k},
\label{axial}
\ende
so that $k\simeq \pi$ describes a  circular cell pattern. As always $k<\pi$  the 
cells prove to be slightly elongated in axial direction. Moreover, the  elongation 
does not depend on the magnetic field amplitude and the size of the gap. Even slow 
rotation has a very weak influence  with the expected tendency to make the cells more flat.

\begin{figure}[hbt]
\epsscale{1.0} 
\plotone{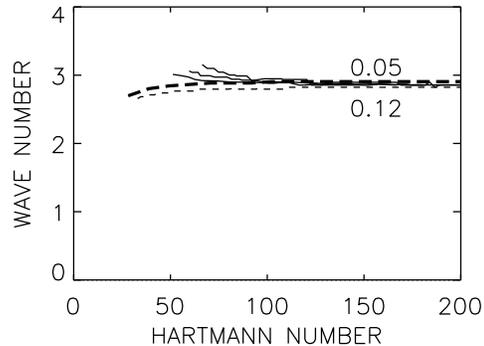} 
\caption{The optimized wave numbers for two values of the radius of the inner cylinder ($r_{\rm in}=0.05$, $r_{\rm in}=0.12$) 
         with perfect-conductor conditions at both boundaries without (dashed) and with rotation of Reynolds 
         numbers 500, 1000 and 2000. The dotted line gives the wave number for circular-shaped cells after (\ref{axial}). The 
         cells are thus elongated in the axial direction by only  8\%. $\rm Pm=10^{-6}$.
} 
\label{wavenumber}
\end{figure}

\section{The experimental  results}

\begin{figure}[hbt]
\epsscale{1.0} 
\plotone{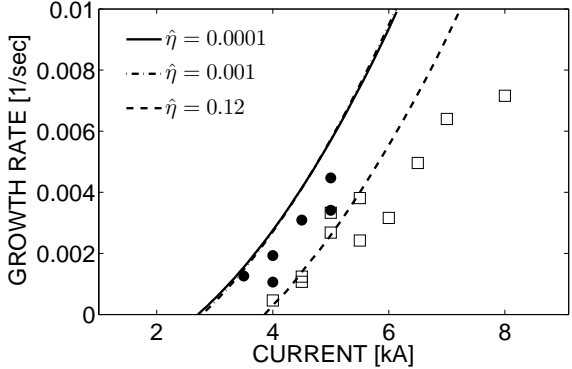} 
\caption{Observed and calculated growth rates in the columnar 
         container with wide gaps: i) inner cylinder with 1.2~cm 
         diameter ($r_{\rm in}=0.12$,  open squares; dashed line) 
         and ii) no inner cylinder (full circles, solid line). In 
         the numerical simulations the inner cylinder is considered 
         as vacuum (dashed line) or as perfect conductor (both left 
         lines). For very small $r_{\rm in}$ the growth rates are 
         independent of $r_{\rm in}$ (dot-dashed line: $r_{\rm in}=0.001$, 
         solid line: $r_{\rm in}=0.0001$).} 
\label{results}
\end{figure}

The experimental apparatus consists of an insulating cylinder with
a height of 75~cm and a radius $R_{\rm out}$ of 5~cm
which is filled with the eutectic alloy GaInSn. An inner cylinder
with radius $R_{\rm in}$ of 0.6~cm can be inserted. At the top and
bottom, the liquid column is in contact with two massive copper
electrodes which are connected by water cooled copper tubes to an
electric power supply. This power supply is able to provide up to
8~kA of current. Although for later experiments the use of
Ultrasonic Doppler Velocimetry (UDV) for the measurement of the
axial velocity perturbation is envisioned, for the first
experiment it has been decided not to use any inserts that could
disturb the homogeneous current from the copper electrodes to the
liquid. With 14 fluxgate sensors the modifications of the magnetic
fields that result from the TI are detected. Eleven of these
sensors are positioned along the vertical axis, while the
remaining three are positioned along the azimuth in the upper
part. Such measurements give the geometry of the field, thus its
shape in azimuthal and axial direction as well as the minimum
current to excite the TI (Seilmayer et al. 2012).

In all cases of instability the observed pattern of the magnetic 
perturbations is a nonaxisymmetric one  with $m=1$. Figure~\ref{results} 
shows the experimental results for the growth rates in comparison
with the theoretical calculations for containers with very wide
gaps between the cylinders. The theoretical result does not depend
on $r_{\rm in}$ for $r_{\rm in}\ll1$ so that these results may serve
as a good proxy for the experiment without any inner cylinder (red
asterisks in Fig.~\ref{results}). For comparison, the data sets of
a second experiment with an inner insulating cylinder with
$r_{\rm in}=0.12$ (green squares in Fig.~\ref{results}) are given.
As expected this experiment requires clearly stronger electrical
currents to excite TI compared to the experiment without an inner
cylinder.

For low growth rates the experimental data fit the
theoretical curves quite well. The agreement is almost perfect for the
container with $r_{\rm in}=0.12$. For  this case one finds a
relation $\omega_{\rm gr}\simeq \gamma (I^2-I_{\rm crit}^2)$ with
$\gamma= 2.7 \cdot 10^{-10}$ so that after Eq.~(\ref{basic}) a
value of $\Gamma\simeq 0.038$ results in perfect agreement with
the theory. The theory, however, always provides the maximum
growth rates optimized with the wave number. The observed growth
rates should thus never lie  {above} the theoretical values
which indeed is the case (except one single square).

\section{Discussion}

Motivated by its astrophysical relevance the Tayler instability has been probed
by an experiment which showed not only its fundamental existence but also many
specific details. To avoid the suppressing action of rotation a resting fluid
conductor for the electric current has been used. We have shown that the
Tayler instability in the experiment appears at the predicted Hartmann
numbers. As expected, the azimuthal Fourier mode with $m=1$ dominates the
unstable magnetic pattern. Also confirmed is the theoretical result that the
critical Hartmann number for nonrotating fluids does not depend on the magnetic
Prandtl number. Finally, the validity of relation (\ref{omg}) for the
growth rate is shown for very small magnetic Prandtl numbers, which has
previously been derived with a linear theory. This relation holds in the
low-conductivity limit $\omega_\eta\ll\Omega_{\rm A}$ while for high
conductivity the growth rate becomes linear in $\Omega_{\rm A}$. The
experiment presented here validates the lower quadratic part of the growth
rate curve; the transition shown in Fig. \ref{rot} towards the linear relation for
high-conductivity, which is characteristic for astrophysical applications,
requires larger Hartmann numbers, thus stronger currents.

For slightly supercritical Hartmann numbers the measured growth rates are in near perfect 
agreement with the theoretical values.
For higher Hartmann numbers, however, they lie  somewhat 
{\em below} the theoretical values. A stabilization effect evidently exists and
is more effective for higher amplitudes of the electrical current than for lower 
values. A natural explanation of this phenomenon could be the generation  
of axial field components by nonaxisymmetric axial flows.  Test calculations 
have shown that indeed even a nonaxisymmetric axial field pattern would reproduce 
the observed reduction of the growth rates for supercritical electric currents. 
Another possible  explanation, at least of parts of this effect, is suggested 
by Eq.~(\ref{omg}) as an 
amplification of the magnetic diffusivity $\eta$ during the experiment. This 
can be due to the internal heating of the fluid conductor which leads to 
higher $\eta$ via its temperature dependence (about 10\% by increase of the 
temperature of 75~K). The counteracting density reduction is much weaker. The 
data in Fig.~\ref{results} do not exclude this possibility. For its final 
confirmation further measurements for stronger currents are necessary. 

Another dynamical stabilization may follow from the production of an extra 
turbulence diffusivity $\eta_{\rm T}$ by  the TI itself.  By numerical
simulations we have shown that the maximal value of the ratio
$\eta_{\rm T}/\eta$ is only of order unity (Gellert \& R\"udiger
2009) which could also  be consistent with the presented data.
Further experiments with  special container constructions to reduce 
the internal heating must show whether the stabilization shown in 
Fig.~\ref{results} is due to the possible axial field component  
or  also due to temperature-effects and dynamical influences.

\begin{figure}[hbt]
\epsscale{1.0} 
\plotone{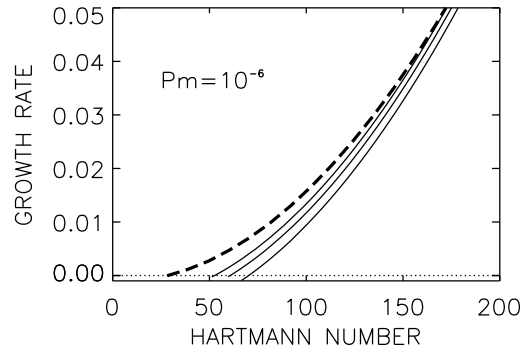} 
\caption{The normalized growth rate of TI in a container with $r_{\rm in}=0.05$ without  
         rotation (dashed) and with rotation of Reynolds numbers 500, 1000 and 2000. 
         The reference frequency is $\sqrt{\nu\eta}/R^2_{\rm out}$. Note the suppression 
         of the instability by the global rotation.}
\label{rot}
\end{figure}

That the radial scale does not appear in the growth rate (\ref{omg}) is an exception. 
It changes for strong fields ($\omega_{\rm gr}\propto \Omega_{\rm A}$) as well as for 
weak fields when taking into account the (suppressing) influence of a given basic 
rotation. Then growth rates also depend on the radial scale. For slow rotation the 
growth rate (\ref{omg}) then formally must be divided by the magnetic Reynolds number 
of this rotation so that $\omega_{\rm gr}\propto \Omega_{\rm A}^2/\Omega$ results. 
The calculations show that this rotation-modified theoretical form of the growth rate 
of the TI can also be probed experimentally. For strong enough magnetic fields, however,
the rotation does no longer influence the linear relation between $\omega_{\rm gr}$  
and $\Omega_{\rm A}$ . Thus experiments with resting containers only slightly overestimate 
the characteristic growth rates, if the rotation is not too fast (Fig. \ref{rot}).

It is obvious that the experiment possesses a great potential for studies about the TI-induced 
eddy diffusivities and -- if the background field is modified by axial field components to helical 
structures -- about the formation and evolution of small-scale kinetic and magnetic helicities and 
even the $\alpha$-effect.

\acknowledgments This work was supported by the Deutsche
Forschungsgemeinschaft (SPP 1488 {\em PlanetMag}) and SFB 609.

\end{document}